# Design Optimization for an Electro-Thermally Actuated Polymeric Microgripper

R. Voicu, R. Muller, L. Eftime

National Institute for R&D in Microtechnologies (IMT-Bucharest)

126A, Erou Iancu Nicolae Street,

077190, Bucharest, ROMANIA

Email: rodica.voicu@imt.ro

*Abstract-* **Thermal micro-actuators are a promising solution to the need for large-displacement, gentle handling force, low-power MEMS actuators. Potential applications of these devices are micro-relays, assembling and miniature medical instrumentation. In this paper the development of thermal microactuators based on SU-8 polymer is described. The paper presents the development of a new microgripper which can realize a movement of the gripping arms with possibility for positioning and manipulating of the gripped object. Two models of polymeric microgripper electro-thermo-mechanical actuated, using low actuation voltages, designed for SU-8 polymer fabrication were presented. The electro-thermal microgrippers were designed and optimized using finite element simulations. Electro-thermo-mechanical simulations based on finite element method were performed for each of the model in order to compare the results. Preliminary experimental tests were carried out**.

## I. INTRODUCTION

Microgrippers are promising tools used as end-effectors for systems that can handle and manipulate micro and nano scaled objects with application in various field of science and industry [1, 2].

In the MEMS area, there are many applications where microgrippers have reached an increasing importance. The applications of microgripper structures range from micro manipulation of micro particles, micro components and even cells to assembling and medical applications, and already have a serious impact on present and future technologies [1-9].

Several microgrippers have been developed using different kind of actuating methods, by means of different physical effects, as piezoelectric, electromagnetic, electrostatic, electro-thermal, etc. Thermal actuation microgripper generally use thermal expansion by Joule heating, generated by an applied voltage. From various actuation principles that can be used in the design of microgrippers, electro-thermal actuators have some advantages as large displacements and relative low response time. The electric current passing through the thin metallic layer generates heats that deform/expand the material of the hot arms, the movement of the arms occurring.

Microgrippers manufactured of polymeric materials offer the advantage of a large displacement and a gentle handling force that can be ideal particularly for specific bio-particle manipulation, as cells. Operating temperatures below $100\ ^0C$ allow the manipulation of living cells and tissues. Polymers offer a much lower Young's modulus, and thus much lower actuation and handling forces [4, 5]. For example, SU-8 has a Young's modulus in the range of 4.02 GPa[6] - 4.95 +/-0.42 GPa [7] and it was intensively used as structural material for microgrippers in the last years. Because of its biocompatible properties, SU-8 polymer can be used in a great variety of bio-MEMS applications [8, 9].

This paper presents two models of polymeric microgripper electro-thermo-mechanical actuated, using low actuation voltages, designed for SU-8 polymer fabrication. The electro-thermal microgrippers consist of two arms configurations 'hot and cold' and were designed and optimized, using finite element simulations [10]. Coupled electro-thermo-mechanical simulations, using COVENTORWARE tools, have been performed in order to describe the microgrippers behaviours in air and liquid as function of the applied voltage.

A comparison between different designs of the microgripper structures and between simulation results was performed in order to optimize the model. We obtained results for different environments operations where the microgripper manipulate the micro particles, as air and liquids. The optimization of the designed models was performed using finite element simulations and numerical methods in order to obtain a stable structure without any out of plane displacements that can occur.

## II. DESIGN AND SIMULATIONS

### A. Design configurations

The designed microgripper structures consist of some major parts: the fixed part, hot and cold underarms, arms and heaters (that we will be calling up and down parts) (see Fig. 1).

The main dimensions of the microgrippers are 200 micrometers X 460 micrometers X 20.6 micrometers. The length of the all free arms is 400 micrometers, 2 micrometers is the thickness of the thermal oxide layer (SiO2), 20 micrometers the thickness of the polymer layer (SU-8), 0.3 micrometers the thickness for each of the Chromium/Gold layers (which are deposited on both sides of the SU-8 layer for the first model) (see Fig. 1).

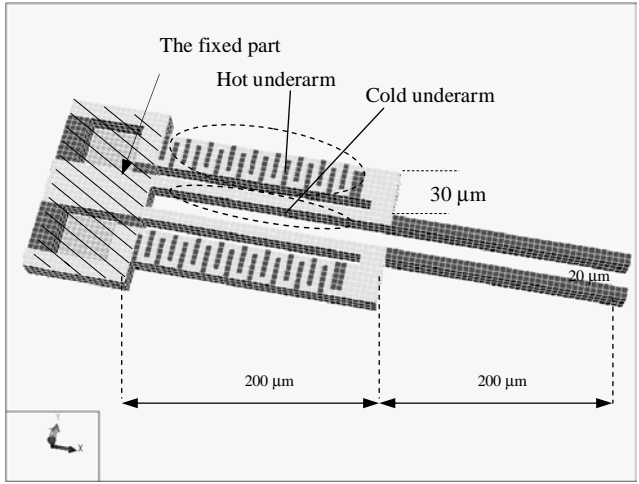

*Fig. 1 3D-meshed model of the microgripper simulated with CoventorWare program*

The second microgripper model has one metallic layer with the same shape and has similar design as the first model with the difference only regarding the position of the metallic layer. So, for the second microgripper model the Chromium/Gold metallic layer with a thickness of 0.3 micrometers is deposited between two SU-8 layers, each with 10 micrometers thickness (see Fig. 2).

In operation, an electrical current will be applied to the gold element. The current will travel through the metallic layers producing the following movement: the part of the polymeric arms, which contains the metallic layers, will be heated and will expand, causing the displacements of the microgripper arms, and occurring the microgripper tips close.

The microgripper structure is initially in the open position with an opening distance of 20 micrometers. The microgripper is designed to operate through an integrated thermal element which is controlled by applying a voltage. The application of a voltage to the metallic layers of the structure will produce the closing of the microgripper and an object can be gripped. When we switch off the voltage, the arms of the structure will open, releasing the micro object.

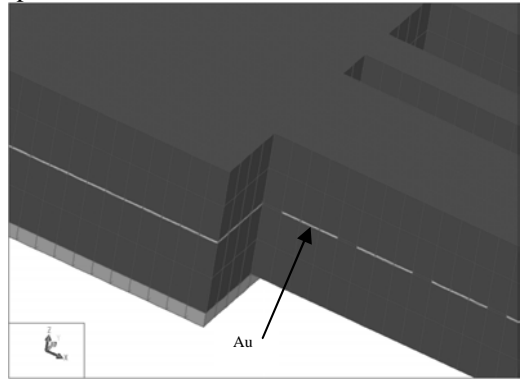

*Fig. 2 3D-detail of the second microgripper model simulated with CoventorWare program*

### B. Simulation results

Coupled electro-thermo-mechanical simulations, using COVENTORWARE tools, have been performed in order to describe the microgrippers behaviours in air and liquid as function of the applied voltage.

The model was meshed using hexahedral elements with 27 nodes and the numbers of volume element was optimised choosing the proper size of the mesh element.

As thermal boundary conditions, necessary to be set for simulations, the initial temperature of the whole structure and the temperature of the environment were considered to be 27 °C and the air convection coefficient was set to 20 W/m²K.

The material properties used in simulations are listed in Table 1.

TABLE 1

MATERIAL PROPERTIES USED IN SIMULATIONS

| Material | SU-8 | SiO2 | Gold |
|---|---|---|---|
| Density (Kg/µm3) X $10^{-15}$ | 1.2 | 2.15 | 19.3 |
| Young's Modulus (MPa) X $10^3$ | 4.95 | 70 | 57 |
| Poisson's Ratio | 0.22 | 0.17 | 0.35 |
| TCE (1/K) X $10^{-5}$ | 5.2 | 0.05 | 1.41 (300K) |
| Thermal conductivity (pW/µmK) X $10^5$ | 2 | 14 | 2970 |
| Specific Heat (pJ/kgK) X $10^{15}$ | 1.675 | 1 | 0.129 |

We have analysed different displacements that can occur in the behaviour of the microgripper models when it is actuated.

Fig. 3 and Fig. 4 show the remaining distance between tips of the microgripper arms and how the arms of the microgripper were closing when we apply 0 - 0.3 Volts on the metallic heaters.

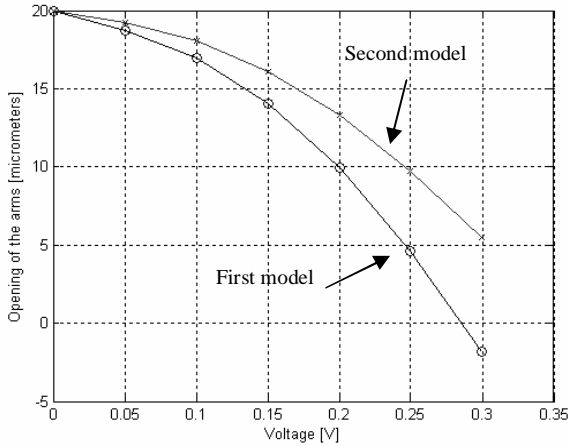

*Fig. 3 The remaining distance between the ends of free arms of the microgripper models*

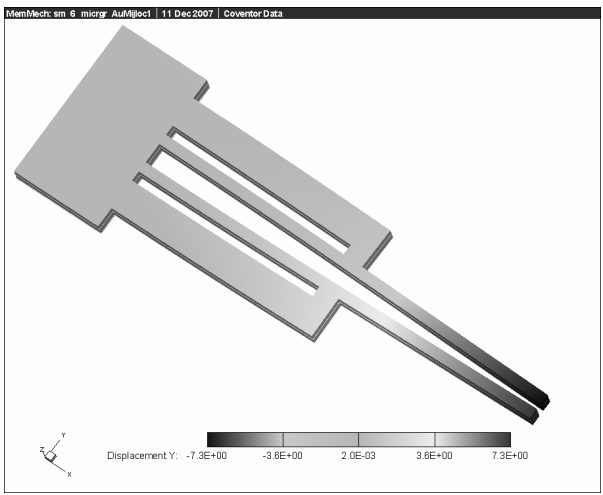

*Fig. 4 The lateral displacements of the second microgripper when 0.25 V is applied – simulation*

It can be observed that for the second microgripper model the arms are closing slower than for the first model for the same values of voltages.

Fig. 5 illustrates the maximum values of the temperature that can occur in the models structures (in the hot arms) when 0-0.3 Volts were applied. For the second model the maximum values of the temperatures were jest a little smaller than the values for the first model. In Fig. 6 it is illustrated the distribution of the temperature in the 3D microgripper model simulated with CoventorWare tools when we apply 0.25 Volts.

Fig. 7 shows the out of plane displacements- the displacement in the vertical direction – for the same voltage applied on the metallic layers. For the second microgripper model the vertical displacements values are smaller with respect to the same voltage values. This means that the second model structure in less instable; for the same voltage values the microgripper is bending down slower (for the 0.05-0.3 Volts the vertical displacement increase with less then one micrometer). Further optimised design of the model can be done varying the dimension of thickness of the polymeric layers and of the metallic layer.

Our aim was to obtain an optimised design of the microgripper model in order to reduce the out of plane displacements that can occur in the microgripper behaviour when is actuated. The Fig. 5 show that the temperatures are similar and the Fig. 7 show an optimised vertical displacement.

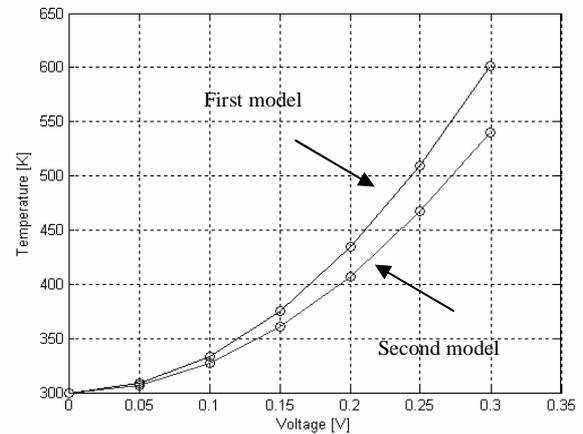

*Fig. 5 Maximal values of the temperatures in the microgrippers when a voltage is applied*

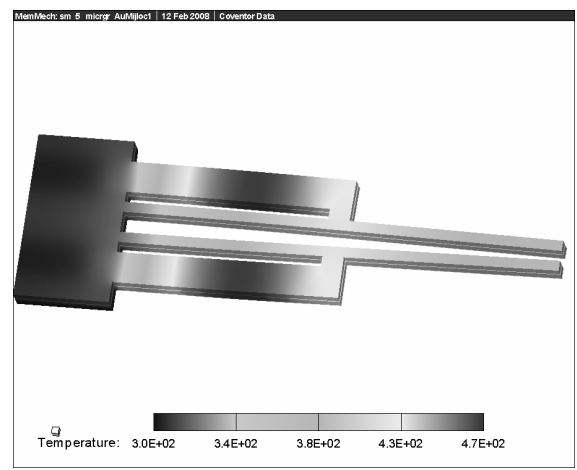

Fig. 6 Simulation of the temperature's distribution in the arms of the microgripper when a voltage of 0.25V is applied (temperature scale in K)

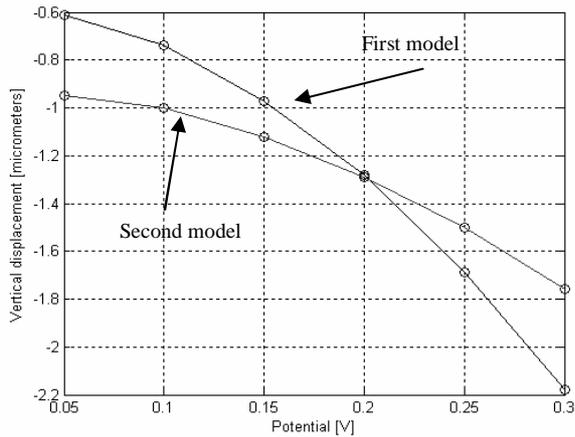

*Fig. 7 Maximal out of plane displacements for the models of microgripper when a voltage is applied*

In order to observe the gripping force that occurred between tips of the first microgripper arms and the micro object gripped we have performed new simulation. In Fig. 8 it can be observed the contact between microgripper tips and the micro object. When we apply 0.25 Volts on the heaters the microgripper model can grippe a micro object with a diameter of 5 micrometers (or 10 micrometers for the second model). For the same voltage value we have simulated the contact pressure (0.061-0.17 MegaPascals) that appear between the tips and micro object (see Fig. 8) and it is obvious that we obtain a small gripping force needed to not damage the sample. These demonstrate the gentle handling force give by the polymeric material used to fabricate the microgripper arms.

To observe the behaviour of the microgripper operating in different liquids we have simulate the first model changing the convection coefficient settings. The values of the convection coefficient for different liquids are higher and variable.

It can be observed that operation in liquid requires higher voltages than operation in air do to the higher convection coefficient encountered in liquids (see Fig. 9).

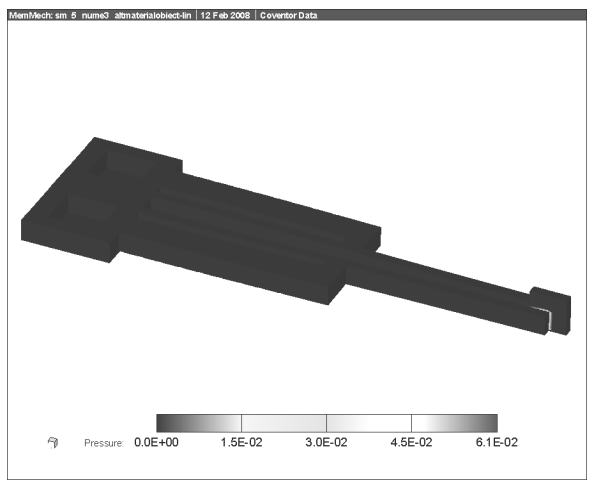

*Fig. 8 Microgripper tips in contact with a micro object gripped - simulation*

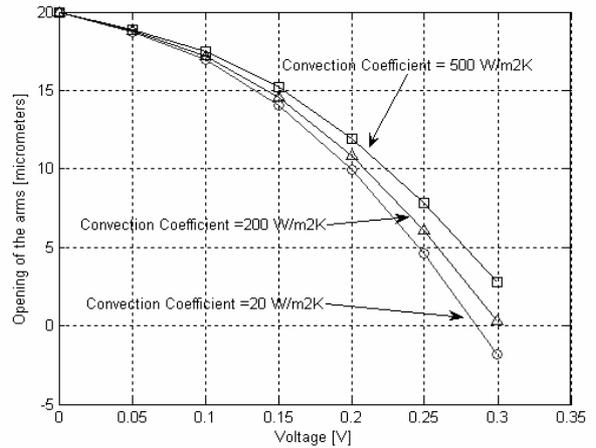

*Fig. 9 The remaining distance between the ends of free arms of the first model of microgripper for different convection coefficient: from air to different aqueous environments*

### III. EXPERIMENTAL RESULTS

Preliminary experimental work was performed for optimize the model of microgripper, using SU-8 polymer as structural material, because of his high thermal expansion coefficient and the possibility to obtain high aspect ration structures.

The fabrication was based on a 4 masks process, based on surface micromachining, using 1.7 micrometers $SiO_2$ as sacrificial layer. The microgripper's arms were manufactured by SU-8 polymer, hard bake at 150 ºC on a hot plate, having a thickness of 20 micrometers; they can move laterally. The heating resistance and the pads were of thin Cr /Au layer. The first steps of the technological process can be seen in Fig. 10.

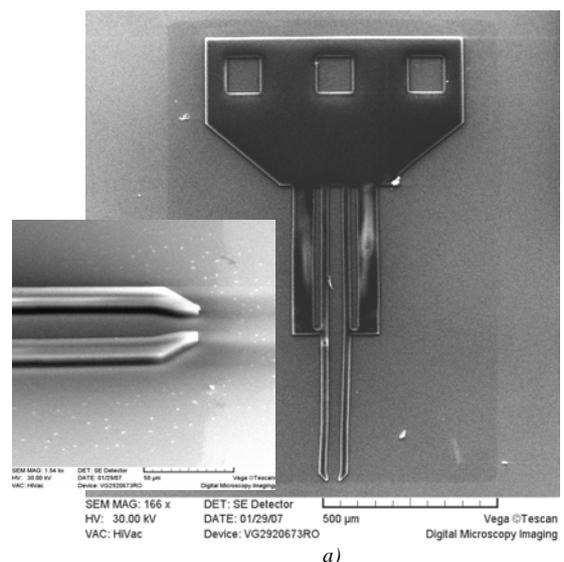

*a)*

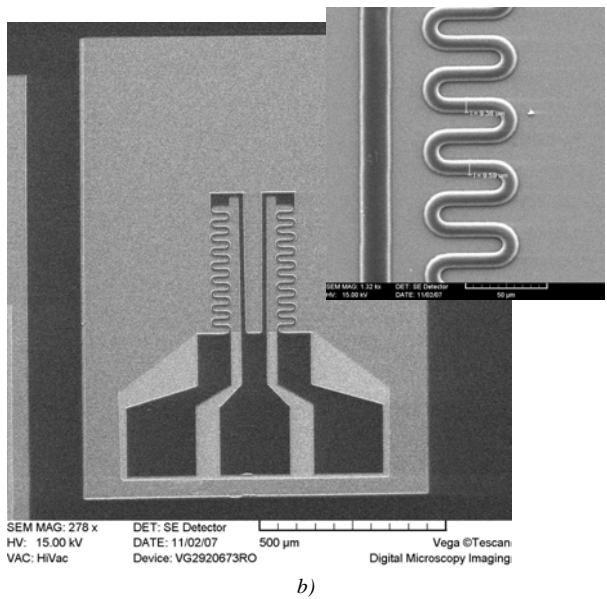

*Fig.10 SEM picture of the (a) SU-8 microgripper (b) Cr / Au resistance fabricated based on FEM model*

## IV. CONCLUSIONS

Electro-thermally actuated microgrippers with a novel configuration for thermal actuators have been designed and characterized and of some extends fabricated and characterized. Preliminary experimental fabrication steps were performed. The maximum temperature reached in the hot underarms exceeds 160 $^{o}$C, but on the free ends of closing arms it is lower then 100 $^{o}$C, making possible even the manipulation of biological samples.

The choice of biocompatible materials together with the low actuation voltages required and large deflection produced at relatively low temperatures makes this microgripper highly suitable for bio-manipulation experiments in air or in aqueous media.

Future work includes the realization of experiments for micro manipulation of micro objects and biological particles.

We have proposed a new model optimized for a micropgripper, electro-thermal actuated at low voltage, manufacturing using polymeric "hot and cold" arms.

ACKNOWLEDGMENT

This work was supported by National Romanian Program INFOSOC-Project Nr. 28/2005 and CALIST- Project Nr.6111/2005.

REFERENCES

[1] K. Kim, E. Nilsen, T. Huang, A. Kim, M. Ellis, G. Skidmore, J.-B. Lee. Metallic microgripper with SU-8 adaptor as end-effectors for heterogeneous micro/nano assembly applications. Microsystems Technologies 10 (2004) 689–693.

[2] J. Kim et al. Polysilicon microgripper. Sensors&Actuators A 33 (1992) 221-7.

[3] S. Butefisch et all., "Novel micro-pneumatic actuator for MEMS", Sensors Actuators A 97-98, 638-645, 2001

[4] N.-T. Nguyen, S.-S. Ho, and C. Lee-Ngo Low, " A polymeric microgripper with integrated thermal actuators", *J. Micromech. Microeng.* 14 (2004) 969–974.

[5] I. Roch et al , "Fabrication and characterisation of an SU-8 gripper actuated by a shape memory alloy thin film", 2003 *J. Micromech. Microeng.*, 13, 330-336

[6] H. Lorenz, M. Despont, M. Fahrni, N. LaBianca, P. Vettiger, and P. Renaud. SU-8: a low-cost negative resist for MEMS. J. Micromech. Microeng 7(1997), 121-124

[7] L. Dellmann, S. Roth, C. Beuret, G. Racine, H. Lorenz, M. Despont, P. Renaud, P. Vettiger, and N. de Rooij. Fabrication process of high aspect ratio elastic structures for piezoelectric motor applications. in Proc. Transducers 1997, Chicago, (1997), 641-644

[8] G. Voskerician, M. S. Shive, R. S. Shawgo, H. von Recum, J. M. Anderson, M. J. Cima, and R. Langer. "Biocompatibility and biofouling of MEMS drug delivery devices", *Biomaterials*, vol. 24, pp. 1959–1967, 2003

[9] N. Chronis, L. P. Lee, "Electrothermally activated SU-8 microgripper for single cell manipulation in solution", *Journal of Microelectromechanical Systems* (2005)

[10] R. Voicu, D. Esinenco, R. Müller, L. Eftime, C. Tibeica, "Method for overcoming the unwanted displacements of an electro-thermally actuated microgripper", *4M Conference 2007, Borovets, Bulgaria*, pp.39-42, oct. 2007